\documentclass[conference]{IEEEtran}
\IEEEoverridecommandlockouts
\usepackage{cite}
\usepackage{amsmath,amssymb,amsfonts}
\usepackage{algorithmic}
\usepackage{graphicx}
\usepackage{dblfloatfix}
\usepackage{pdfpages}
\usepackage{textcomp}
\usepackage{caption}
\usepackage{balance}
\usepackage{url}
\usepackage{xcolor}
\def\BibTeX{{\rm B\kern-.05em{\sc i\kern-.025em b}\kern-.08em
    T\kern-.1667em\lower.7ex\hbox{E}\kern-.125emX}}
\begin{document}

\title{Analysis of Superconducting Qubit Layouts Using InductEx\\
\thanks{This work was supported under the NIWC-PAC In-House Innovation Program.}
}

\author{\IEEEauthorblockN{Sean Crowe\textsuperscript{1}}
\IEEEauthorblockA{\textit{CERF Laboratory} \\ 
\textit{NIWC-PAC}\\
San Diego, United States \\ 
sean.t.crowe2.civ@us.navy.mil}
\and
\and
\IEEEauthorblockN{Benjamin Taylor}
\IEEEauthorblockA{\textit{CERF Laboratory} \\ 
\textit{NIWC-PAC}\\
San Diego, United States \\
benjamin.j.taylor10.civ@us.navy.mil}
\and
\IEEEauthorblockN{Nicholas Ferrante}
\IEEEauthorblockA{\textit{CERF Laboratory} \\ 
\textit{NIWC-PAC}\\
San Diego, United States \\
nicholas.b.ferrante.civ@us.navy.mil}
\and
\IEEEauthorblockN{\hspace{1cm} Brad Liu}
\IEEEauthorblockA{\hspace{1cm} \textit{CERF Laboratory} \\ 
\hspace{1cm} \textit{NIWC-PAC}\\
\hspace{1cm} San Diego, United States \\ 
\hspace{1cm} brad.c.liu.civ@us.navy.mil}
\and
\IEEEauthorblockN{\hspace{-0.9cm}Susan Berggren}
\IEEEauthorblockA{\hspace{-0.9cm}\textit{CERF Laboratory} \\ 
\hspace{-0.9cm}\textit{NIWC-PAC}\\
\hspace{-0.9cm}San Diego, United States \\ 
\hspace{-0.9cm}susan.a.berggren.civ@us.navy.mil}
\and


}

\maketitle

\begin{abstract}

InductEx \cite{InductEx} is a software tool used for the analysis of integrated circuit designs and extraction of design parameters by way of numerical electromagnetic field solving. This tool was originally developed with Rapid Single Flux Quantum (RSFQ) chips in mind, but it has a broad applicability and can be extended to other processes. In this poster, we report a comprehensive analysis of a superconducting aluminum two qubit chip. This analysis was performed with InductEx.

We report the design of a two qubit chip which has the characteristics necessary to execute single and two qubit gates. Ahead of fabrication, several design characteristics have been extracted from this quantum chip design in order to verify that it satisfies basic design principles of transmon qubits. These characteristics are reported in this poster and they include the calculation of chip anharmonicities, qubit frequencies, resonator frequencies as well as g-factors and dispersive shifts. Design constraints which are satisfied by these extracted parameters are discussed.  Additionally, qualitative aspects of the chip have been obtained from current density maps and are reported here. 

Taken as a whole, this analysis demonstrates the broad applicability of Inductex to integrated circuit design and particularly to the problem of quantum circuit layout optimization.

\end{abstract}




\section{Extended Poster Abstract}


When designing a superconducting qubit chip, one typically starts with a circuit schematic of the chip, where the qubits are realized as capacitively shunted SQUIDs which are coupled with a resonator which in turn is coupled with a readout line. Circuit parameters are then chosen in order to satisfy certain design constraints related to chip performance \cite{Qengineering}. Creating a circuit layout which is faithful to the schematic is a non-trivial problem which requires special care.

\footnotetext[1]{Full Affiliation: Cryogenic Exploitation of Radio Frequency Laboratory, \\ Naval Information Warfare Center Pacific}

\begin{figure}
    \centering
    \includegraphics[width=0.75\linewidth]{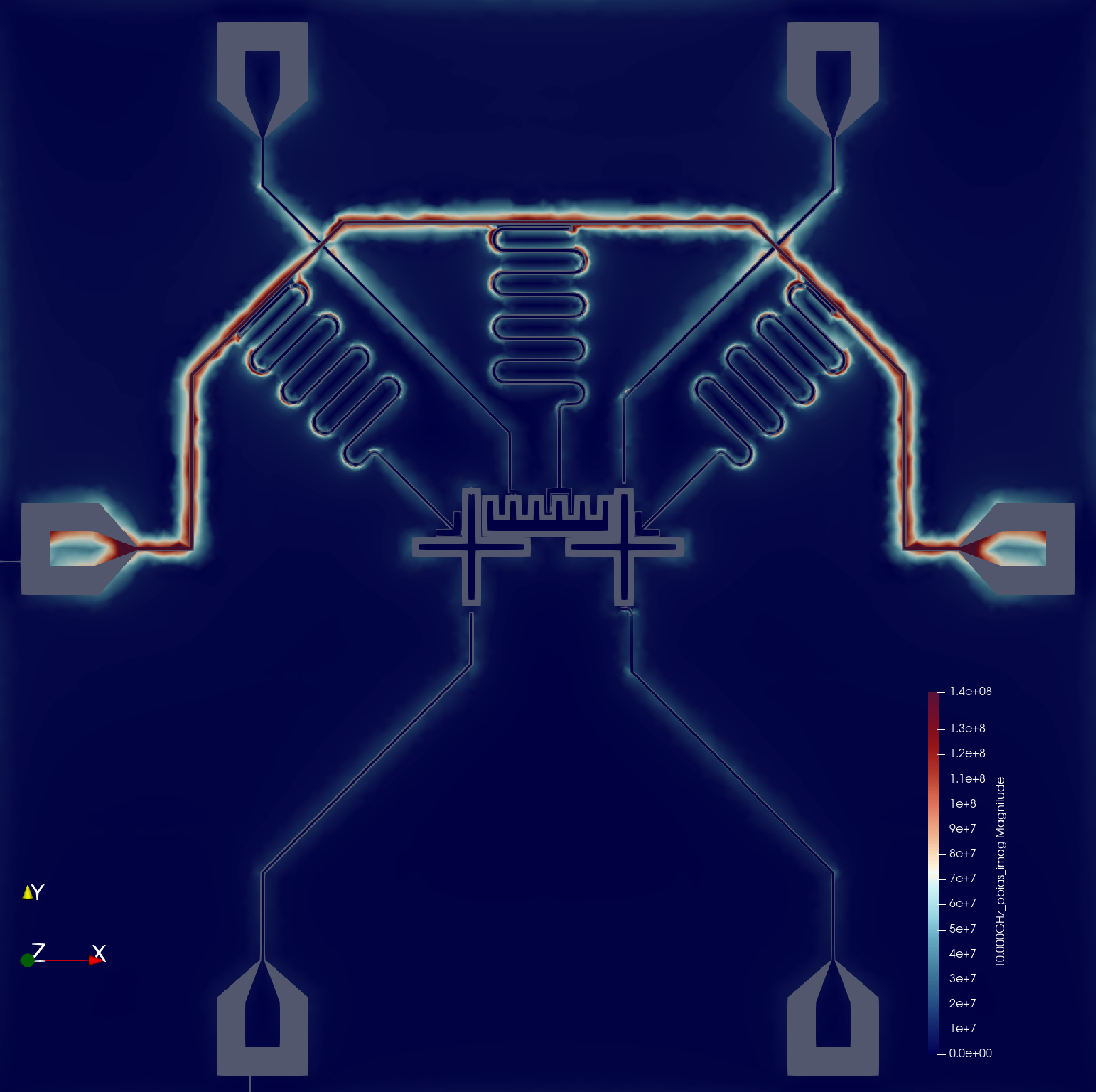}
    \caption{Simulation of a readout operation on our two qubit chip. This readout operation was performed with InductEx and The color map is based on the magnitude of the current density vector field. The driving frequency is $7\, {\rm \ GHz}$.}
    \label{fig:readout_line_sim}
\end{figure}

In this poster, we detail the optimization of the quantum chip layout shown in Fig:~\ref{fig:readout_line_sim}. This chip layout consists of two transmon qubits and a tunable coupler. The coupler design was primarily motivated by \cite{tunable}. It was created using the open source program KLayout, and Inductex was used to perform the extraction of the circuit parameters.

Qubit and coupler capacitances were determined by using Inductex to calculate the capacitance matrix between all pieces of metal on the device layer of the chip. The capacitance between the first qubit and the ground was found to be $C_{q1}=108 {\rm \ fF}$. Critical currents of the Josephson junctions in the qubits were determined based on the junction area and the the critical current density of the fabrication process. The first qubit is a fixed frequency qubit with only one junction. The critical current of this junction is $I_{c1}=30.0 {\rm \ nA}$. Taken together, these figures imply the excitation energy $E^{(1)}_{01}=4.43 {\rm \ GHz}$, and an anharmonicity $\alpha_{1}=198 {\rm \ MHz}$. The anharmonicity was computed using numerical diagonalization of the Hamiltonian. The ratio of the charging energy of the junction to the charging energy of the capacitor was found to be $E_{j}^{(1)}/E_{C}^{(1)}=83.1$.

The second qubit is a flux biased SQUID shunted by a capacitor. It is therefore tunable. The critical current of the SQUID is $I_{c2}=40.0 {\rm \ nA}$, and the ratio of its junction widths is $W_1/W_2=3$. The shunting capacitor's value was found numerically to be $C_{q2}=108 {\rm \ fF}$. The excitation energy of this qubit at zero biasing field is therefore $E^{(2)}_{01}(0)=5.16 {\rm \ GHz}$ with an anharmonicity of $\alpha_{2}=196 {\rm \ MHz}$. Because of the asymmetry of this transmon, the lowest reachable excitation energy is $E^{(2)}_{01}\left(\frac{1}{2}\Phi_0\right)=2.58 {\rm \ GHz}$, which occurs when half of a flux quantum is biasing the SQUID loop. The ratio of charging energies for this qubit was found to be $E_{j}^{(2)}/E_{C}^{(2)}=111$.

The tunable coupler is also a flux biased SQUID shunted by a capacitor. The value of its capacitance was found to be $C_{tc}=126 {\rm \ fF}$. The critical current of its junction is $I_{ct}=35.0 {\rm \ nA}$. The ratio of its junction widths is also $W_1/W_2=3$. Based on the capacitance and the critical current, the excitation energy of the tunable coupler is $E^{(t)}_{01}(0)=4.47 {\rm \ GHz}$ with an anharmonicity of $\alpha_t=196 {\rm \ MHz}$. The lowest reachable excitation energy for the coupler occurs when half of a flux quantum is biasing the SQUID loop and this energy is  $E^{(t)}_{01}\left(\frac{1}{2}\Phi_0\right)=2.24 {\rm \ GHz}$.  The charging ratio for the coupler is $E_{j}^{(t)}/E_{C}^{(t)}=113$.

In addition to qubit frequencies, resonator frequencies were also calculated. This calculation was done by treating the the resonator as an harmonic oscillator and then extracting the inductance and capacitance of the resonator. 

The inductance which was extracted for the first resonator was found to be $L_{r1}=1.96 {\rm \ nH}$. The extracted capacitance for this device was found to be $C_{r1}=744 {\rm \ fF}$. This implies a fundamental resonant frequency of $f_{r1}=6.55 {\rm \ GHz}$. An analytic estimate can be obtained from the formula $f_{r1}=\frac{c}{4 l \sqrt{\varepsilon_{\rm eff}}}$, where l is the length of the resonator and $\varepsilon_{\rm eff}$ is the effective dielectric constant of the resonator. Because the fabrication process uses a thick substrate, we can approximate $\varepsilon_{\rm eff}\approx\frac{1}{2}\left(\varepsilon_{\rm substrate}+1\right)=6.2$ for silicon \cite{CPW}. The length of the resonator was determined to be $4320 \mu{\rm m}$ by numerical integration. Taken together, these imply an analytic approximation for the frequency to be $f_{r1}^{\rm analytic}=6.96 {\rm \ GHz}$. This is close to the numerical value obtained by InductEx, however, this analytic approximation does not account for the detailed geometry of the resonator and coupler. We therefore consider the numerically obtained frequency more reliable.

Given the characteristics of the resonator, it is possible to obtain estimates for the g-factor of the first qubit and also of the dispersive shift. The g factor is given by $g_{q1-r1}=2 e \beta V_{\rm rms}\langle 0|\hat{n}|1\rangle$ , where $\beta=\frac{C_{q1-r1}}{C_{q1-r1}+C_{q1}}$, $V_{\rm rms}=\sqrt{\frac{h f_{\rm r1}}{2 C_{r1}}}$, and $\hat{n}$ is an operator which measures the number of cooper pairs on the capacitor \cite{basic}. $C_{q1-r1}$ is the capacitive coupling between the resonator and the first qubit. Numerically this value was found to be $C_{q1-r1}=7.59 {\rm \ fF}$. In units where $h=1$, this works out to be $g_{q1-r1}=67.9 {\rm \ MHz}$. Finally, the dispersive shift is given by $\chi_{\rm q1}=347 {\rm \ kHz}$.

The second resonator couples directly to the tunable coupler. Inductex was used to calculate both the inductance and capacitance of this second resonator. The value of inductance was found to be $L_{r2}=1.99 {\rm \ nH}$. The value of the capacitance was found to be $C_{rt}=740 {\rm \ fF}$. This implies a fundamental frequency for the second resonator of $f_{\rm rt}=6.51 {\rm \ GHz}$. Additionally, a capactive coupling between the second readout resonator and the tunable coupler is found to be $C_{rt-t}=6.54 {\rm \ fF}$. Again, taken together, this implies the g-factor $g_{rt-t}=54.7 {\rm \ MHz}$ and a dispersive shift of $\chi_{t}=226 {\rm \ kHz}$.

The third resonator couples to the second qubit. The inductance of this resonator was found to be $L_{r3}=1.95 {\rm \ nH}$. The value of the capacitance for this resonator was found to be $C_{\rm r3}=722 {\rm \ fF}$. This implies a fundamental frequency of $f_{\rm r3}=6.66 {\rm \ GHz}$. The capacitive coupling between the the third readout resonator and the second qubit was found to be $C_{\rm r3-q2}=7.72 {\rm \ fF}$. Given this information, the g-factor between the second qubit and the resonator was found to be $g_{\rm r3-q2}=75.7 {\rm \ MHz}$ and the dispersive shift was found to be $\chi_{q2}=607 {\rm \ kHz}$.


Finally, the capacitive coupling between the two qubits and their respective XY drive lines was computed. The coupling between the first qubit and its XY drive line was found to be $C_{\rm d1-q1}=0.201 {\rm \ fF}$. The coupling between the second qubit and its drive line was found to be $C_{d2-q2}=0.194 {\rm \ fF}$. 

In conclusion, we have used InductEx in order to perform a detailed analysis of a two qubit superconducting chip. This analysis included extraction of qubit frequencies, resonator frequencies, and anharmonicities. Extraction of resonator capacitances allowed for predictions of dispersive shifts and coupling strengths between qubits and resonators. Moreover, extraction of coupling strengths between the qubits and the coupler allow for expectations of the characteristics of two qubit gates ahead of fabrication. Taken as a whole, this analysis demonstrates InductEx is a robust tool which is applicable to the problem of superconducting qubit layout optimization.


\balance


\end{document}